\documentclass[preprint,5p]{elsarticle}

\usepackage{xcolor}
\usepackage{graphicx}
\usepackage{lineno,hyperref}
\usepackage{amsmath}
\usepackage{amssymb}
\usepackage{subcaption}
\modulolinenumbers[10]

\begin{document}

\begin{frontmatter}

\title{Coupling Fluid Plasma and Kinetic Neutral Models using Correlated Monte Carlo Methods}

\author{G.J. Parker, M.V. Umansky, B.D. Dudson}

\address{Stanford University, Department of Mathematics, Stanford CA 94305, USA}
\address{Lawrence Livermore National Lab, Livermore, CA 94550, USA}

\begin{abstract}
While boundary plasmas in present-day tokamaks generally fall in a fluid regime, neutral species near the boundary often require kinetic models due to long mean-free-paths compared to characteristic spatial scales in the region. Monte-Carlo (MC) methods provide a complete, high-fidelity approach to solving kinetic models, and must be coupled to fluid plasma models to simulate the full plasma-neutrals system. The statistical nature of MC methods, however, prevents the convergence of coupled fluid-kinetic simulations to an exact self-consistent steady-state. Moreover, this forces the use of explicit methods that can suffer from numerical errors and require huge computational resources. 

Correlated Monte-Carlo (CMC) methods are expected to alleviate these issues but have historically enjoyed only mixed success. Here, a fully implicit method for 
coupled plasma-neutral systems is demonstrated in 1D using the UEDGE plasma code and a homemade CMC code. In particular, it is shown that ensuring the CMC method is a differentiable function of the background plasma is sufficient to employ a Jacobian-Free Newton-Krylov solver for implicit time steps. The convergence of the implicit coupling method is explored and compared with explicit coupling and uncorrelated methods. It is shown that ensuring differentiability by controlling random seeds in the MC is sufficient to achieve convergence, and that the use of implicit time-stepping methods has the potential for improved stability and runtimes over explicit coupling methods.

\end{abstract}


\end{frontmatter}

\section{Introduction}

High-fidelity modeling for coupled systems of plasma and neutral species is essential in the design and study of divertor regions in fusion devices, where strong interactions occur \cite{EdgeFusionDevices,militello2022boundary}. Fluid models provide accurate approximations for the plasma behavior, and robust approaches using finite-volume methods have now been employed with great success for several decades \cite{wiesen2015new}. In cases where the mean-free paths of neutral particles in edge plasmas are long, neutral species exhibit non-fluid-like behavior and are more accurately modeled using the kinetic Boltzmann equation \cite{van2022assessment,blommaert2019spatially}. Monte-Carlo (MC) methods \cite{ EdgeFusionDevices,heifetz1986neutral, EIRENEmanual, DEGASexample} are by far the most developed among the several known approaches \cite{hakim2020conservative, wersal2017neutral} to solving the kinetic equation. 

The stochastic nature of Monte-Carlo methods, however, results in unfavorable convergence properties when coupling to fluid plasma models. Simulations typically converge to stochastic equilibria, for which robust convergence criteria may be difficult to determine, and which are known to exhibit bias compared to true solutions \cite{EdgeFusionDevices,ghoos2016,boeyaert2022numerical}. More importantly, the stochastic nature of Monte-Carlo methods typically forces the use of explicit coupling methods. By standard numerical stability theory  \cite{leveque2007finite, roache1998fundamentals}, such explicit methods can result in large numerical errors and require excessively small timesteps to achieve the desired accuracy.  Moreover, both experiments and modeling indicate the presence of slow transient effects ( $\sim$ 100ms -- 1s) in the divertor detachment front \cite{krasheninnikov1999stability,harrison2017detachment}. As a result, resolving long timescale effects to necessary accuracies using explicit methods can require prohibitively long runtimes or heavy computational resources. Implicit methods enjoy much more desirable stability properties and {  have the potential to advance toward equilibrium with timesteps many orders of magnitude larger than the stability threshold of explicit methods. Using implicit methods for a continuum fluid model coupled with a standard Monte-Carlo calculation, however, would face significant difficulties due to the statistical noise entering the Jacobian evaluation}.

It has been known to experts for many years that correlated Monte-Carlo (CMC) methods have the potential to address these problems, but attempts to realize this have had only mixed success \cite{stotler2000coupling,gerhauser1988modelling,maddison1994recycling}. To the authors' knowledge, correlated Monte-Carlo methods for fusion boundary plasma modeling have only been successfully demonstrated in one spatial dimension, and only using explicit coupling for which the advantages of correlation were shown to be negligible \cite{ghoos2016}. The purpose of this study is to demonstrate a fully implicit coupling scheme in one spatial dimension with robust stability and convergence properties employing correlated methods.  

Section 2 explains the coupled model studied in detail, and reviews the difference between explicit and implicit coupling methods. Section 3 offers a new perspective on Correlated Monte-Carlo methods that is useful in the development of implicit coupling algorithms. Section 4 introduces the UEDGE plasma code, a homemade Monte-Carlo code using the Julia programming language and the methods of coupling these. Section 5 displays the results of coupled simulations, and Sections 6 and 7 provide the discussion and conclusion respectively.

\section{Numerical Methods}

The model investigated couples a set of fluid-plasma evolution equations to a chosen equation modeling neutral species with two basic interactions: (i) ionization, and (ii) charge exchange.   
Let $p(x)$ and $n(x)$ denote the plasma and neutral state vectors, so that e.g. $p=(n_i, n_e, T_i, T_e, u_i)$ is a vector-valued function combining the ion and electron densities, temperatures, and velocities.  
The coupled model leads to a system of equations with the schematic form 
\begin{eqnarray}
\partial _tp+P(p)&=& S_1(p,n)\label{plasma}\\
N(n) & =& S_2(p)\label{neutral}
\end{eqnarray}
\noindent wherein $P,N$ are the plasma and neutral evolution equations respectively, and $S_i$ comprise the interaction terms. {\it We make the additional assumption that the neutral species' relaxation time is on a significantly faster timescale than the plasma timescale} -- in this situation, it is only necessary to consider the time-independent version of $N(n)$. It is additionally assumed that recombination is negligible, and that the source of neutrals from the wall is independent of the neutral distribution, i.e. $S_2$ is independent of $n$.  

While we focus here on two specific plasma models encapsulated by a simple non-linear diffusion equation and the UEDGE model \cite{Rognlien1992, Rognlien1999}, we emphasize that the techniques described are equally applicable to any chosen pairs of plasma and neutral models, or more generally to any chosen pair of continuum (e.g. finite volume) and particle-based (e.g. Monte Carlo) simulations. In particular, including additional physics effects such as recombination of ions, additional neutral species, or { more sophisticated wall interactions} should constitute no fundamental change.  

\subsection{Explicit vs. Implicit Coupling} 

Suppose that known techniques may be applied to time evolve the plasma for a given fixed neutral distribution, and vice versa. That is, assume that one has known numerical algorithms implementing the time evolution operator $U_{\Delta t}(p,S)$ for plasma with timestep $\Delta t$ and fixed source term $S$, and $N^{-1}$ the inverse operator of the time-independent neutral equation. The system (\ref{plasma}--\ref{neutral}) may then be coupled using either explicit or implicit methods. Note that this terminology refers to the method used to couple the system, while the individual solvers for $U_{\Delta t}, N^{-1}$ may themselves include an internal choice of an explicit or implicit method. 

\subsubsection{Explicit Coupling}
Given a coupled state $(p_j, n_j)$ at time $t_j=j\Delta t$ for $j=0,1,...$, an explicit method advances the state by setting 
\begin{eqnarray}
p_{j+1}&=& U_{\Delta t}(p_j  \ , \ S_1(p_j, n_j)) \label{explicitp}\\
n_{j+1} &=& N^{-1}(S_2(p_{j+1} )) \label{explicitn}.
\end{eqnarray}
\noindent The method is explicit in the sense that the states at timestep $j+1$ are defined entirely using quantities calculated in the previous steps. The above demonstrates a standard first-order explicit splitting method, though many more sophisticated variations of explicit coupling are possible { including higher-order or Strang splitting methods, and hybrid implicit-explicit schemes involving dual-time stepping (see below)}. The above time-step is iterated until a sufficient equilibrium is reached. {  In our implementation below, the time evolution operator $U_{\Delta t}$ relies on an internal choice of a first-order implicit Euler method}.  

Explicit coupling methods have several key advantages and disadvantages. Advantages include that they are relatively easy to implement and parallelize, and can be implemented treating the existing individual solvers or codebases for $U_{\Delta t}, N^{-1}$ as ``black boxes''. One main disadvantage is that explicit coupling methods often have quite stringent stability requirements \cite[Ch. 7]{leveque2007finite}. Time-steps must be taken smaller than some stability threshold (analogous to e.g. the CFL condition \cite[Ch. 10.7]{leveque2007finite}) to avoid severe numerical instability. This stability threshold can be many orders of magnitude smaller than the timescales of certain physical effects, making the computations in such cases extremely inefficient; moreover, the stability threshold can often only be determined experimentally and is dynamic as the plasma evolves. Related to that, the second disadvantage of explicit coupling schemes is that they rely on the assumption that the coupling terms are relatively weak, and so for strongly coupled systems convergence is not guaranteed. 
Despite these significant shortcomings, explicit coupling methods are used in boundary plasma modeling for self-consistent calculations using fluid plasma and Monte-Carlo based kinetic neutrals; in fact, {\it explicit coupling is the only approach that currently exists for this kind of calculation}.
Contrasting the above with implicit methods makes the reason for this clear. 

\subsubsection{Implicit Coupling} A first-order implicit method advances the state by $\Delta t$ by taking $(p_{j+1}, n_{j+1})$ as the unique solution of the non-linear system:
\begin{eqnarray}
\hspace{-1.0cm} & &(I +\Delta t P)p_{j+1} - \Delta t  S_1(p_{j+1}, n_{j+1})= p_j\label{implicitplasma} \\ [1mm] 
 & & n_{j+1}= N^{-1}(S_2(p_{j+1}), \label{implicitneutral}
\end{eqnarray}

\noindent where $P$ now denotes the time-independent plasma equation. Substituting (\ref{implicitneutral}) into (\ref{implicitplasma}) results in an equation purely for $p_{j+1}$ defining the evolution of the plasma. For $\Delta t$ sufficiently small, the equation is well-conditioned and may be solved using a suitable implementation of Newton's method, { which is here called a {\it fully implicit }method. The non-linear system (\ref{implicitplasma})--(\ref{implicitneutral}) for each time-step can also be solved by other methods, in particular by dual time-stepping methods \cite{Ghoos2,dualstepping}. When the dual time-steps are advanced by an explicit method, the latter scheme is referred to here as a {\it hybrid implicit--explicit} method.}  

{ Fully implicit methods enjoy far better stability properties than explicit ones, while hybrid implicit--explicit methods enjoy some but not all of these advantages. For fully implicit methods, the time-step may usually in practice }be taken as large as the Newton solver will allow without impacting accuracy  \cite[Ch. 7]{leveque2007finite}. The seemingly innocuous disadvantage of fully implicit methods is that they require the functions $N, P, S_i$ to be differentiable functions of the plasma and neutral states. This is the requirement that precludes the use of Monte-Carlo methods (which are {\it a priori} not even constant for repeated evaluation at the same input). From a mathematical perspective, {\it the key result of this study is to demonstrate that correlated Monte Carlo methods can be made sufficiently differentiable as a function of parameters to apply Newton solvers}, and thus fully mplicit methods.

\section{A New View of Correlated MC}
When the model $N(n)$ for the neutral species is the kinetic Boltzmann equation, Monte-Carlo methods are the most widely-used technique for finding the equilibrium distribution. The standard viewpoint of a Monte-Carlo method (see e.g. \cite{heifetz1986neutral, EIRENEmanual, DEGASexample, EdgeFusionDevices}) is the following: particle test flights are run using (pseudo) random numbers to generate particle trajectories, and each test flight gives a sample from the equilibrium particle distribution. Thus with sufficiently many flights, the tallied output converges to the solution of $N(n)= S_{2}(p)$. 

A correlated Monte-Carlo method does the same, but fixes the random numbers generated for each individual test flight to make the output reproducible (cf. Section 5 of \cite{stotler2000coupling}). The result is a sample from the same equilibrium distribution, and the distribution is now obtained by averaging over increasingly large numbers of possibilities for the fixed random seeds. 

\subsection{Correlated MC Constructs the Inverse Operator}
When considering a Monte-Carlo method as a particular step in a larger algorithm as in (\ref{implicitplasma}--\ref{implicitneutral}), it is advantageous to adopt a more abstract viewpoint. Rather than working at the level of equations as explained above, let us consider the situation at the level of operators. Suppose that the neutral state is described by a vector $n(x)\in \mathbb R^m$ on a discretized spatial grid of size $m$. Thus the Boltzmann equation $N_p(n)=N(n)-S_2(p)$ may be considered as a matrix $N_p\in M(\mathbb R^m , \mathbb R^m)$ in the space of operators on the state space. In our case, this operator is linear, but the same framework applies equally well to a space of nonlinear operators including neutral-neutral interactions.   

The existence of a physical equilibrium distribution dictates that the equation should always be solvable for reasonable plasma backgrounds $p$, i.e. that $N^{-1}_p\in M(\mathbb R^m , \mathbb R^m)$ exists. One may combine all the random choices required for Monte-Carlo flights into a joint probability distribution $\mathbb X$. Then, at an operator level, a correlated Monte-Carlo method is a (parameterized) operator-valued probability distribution 
\begin{eqnarray}P\times \mathbb X& \longrightarrow & M(\mathbb R^m, \mathbb R^m) \\
(p,\xi) & \mapsto& N_{p,\xi}^{-1}.
\end{eqnarray}

\noindent  That associates to any pair of a background plasma state $p$ and joint draw of pseudorandom numbers $\xi$ a particular approximate inverse to $N_p$. The true inverse is then obtained as $N_p^{-1}=\int N^{-1}_{p,\xi} d\mathbb X(\xi)$, i.e. averaging over increasingly large samples of the joint probability distribution $\mathbb X$ gives better approximations to the true inverse. For algorithms that require multiple applications of the operator $N^{-1}_p$ (e.g. explicit or implicit coupling), it can be advantageous to fix a particular choice and carry out the algorithm to completion, rather than generate a new inverse each time $N_p^{-1}$ is called within the algorithm.

 \subsubsection{Smooth Dependence of MC}
 \label{section3.1.1}
 The space $M(\mathbb R^m , \mathbb R^m)$ where the approximate inverse operators are valued is a vector space (more generally a manifold for non-linear operators), as is the space of plasma parameters $p$. It therefore makes sense to ask whether, for fixed $\xi \in \mathbb X$, the operator inverse $N_{p,\xi}^{-1}$ is a continuous, differentiable, or smooth function of the plasma state $p$.  Note that without correlation, continuity certainly fails --- even for fixed $p$ each call to an uncorrelated MC method produces a different result. If differentiability is correctly imposed for fixed $\xi$, this function may be called repeatedly as a step in larger algorithms relying on the linearized approximation of a function, in particular various versions of Newton's method.

 In practice, there are at least two challenges in making a MC output (i.e. $N_{p,\xi}^{-1}$) depend differentiably on $p$. These include:
 \begin{enumerate}
 \item[(i)] De-synchronization of trajectories 
 \vspace{2mm}
 \item[(ii)] Discreteness of estimators 
 \end{enumerate}

\noindent  If one naively fixes the random seeds, particle flights can de-synchronize for infinitesimal changes in background parameters. To elaborate, the cut-off determining whether a particular collision is a charge-exchange or ionization is a discrete transition; thus a tiny variation in background parameters causing a fixed random seed to land on the other side of this cut-off may cause the particle flight to proceed for more steps, thereby shifting the seeds for subsequent flights. Even if this shift is remedied, the estimator used to tally the output will change by a discrete transition when the particle flight elongates, which creates a point of discontinuity with respect to the parameters. 

To fix the first of these issues, it is necessary to identify the joint probability distribution $\mathbb X$ and have a reliable method of generating joint draws so that these shifts do not occur, as well as imposing continuity of the tallies. In practice, the former means pre-generating random numbers for all possible branchings of the particle trajectory before the flight begins. To fix the second, one must use more advanced estimators with better continuity properties. In this study, the attenuated absorption method \cite{heifetz1986neutral} was used with a track length estimator \cite[Eq. 2.35]{DEGASmanual}. If one prefers a collisional estimator, then finite-size particle methods provide continuity properties. {\it In general, it is advantageous to identify any source of discreteness within the MC simulation and replace them with continuous quantities} .

 \begin{figure}[h!]
    \centering

    \begin{subfigure}[b]{0.4\textwidth}
        \caption{}\includegraphics[width=\textwidth]{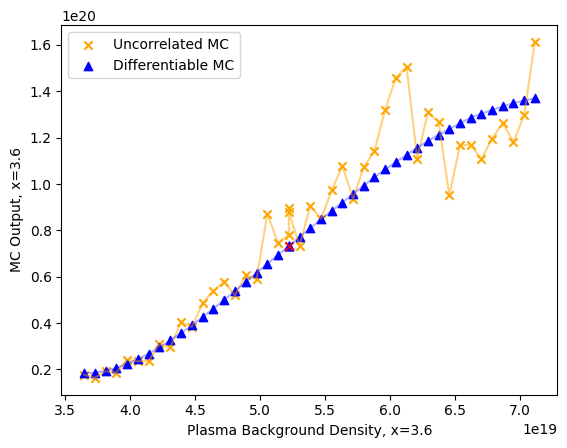}
        
        \label{fig:a}
    \end{subfigure}

    \begin{subfigure}[b]{0.4\textwidth}
        \caption{}\includegraphics[width=\textwidth]{ 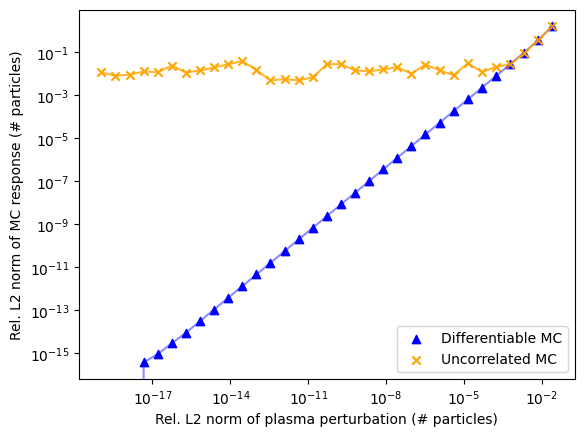}
        
        \label{fig:b}
    \end{subfigure}

    \caption{ The differentiability of the correlated Monte-Carlo method (blue) compared to the discontinuous uncorrelated method (orange). (a, top) The partial derivative (at an equilibrium plasma state with value indicated by the red cross) of a single volume cell with respect to varying the plasma in the direction of a Gaussian perturbation $\Delta p$. (b, bottom) The dependence of the relative $L^2$-norm of the neutral response on the relative $L^2$-norm of the perturbation $s\Delta p$ on a log-log scale.}
   \label{differentiable}
\end{figure}

{ When smoothness is imposed correctly, the directional derivative at a plasma background $p$ in the direction of a perturbation $\Delta p$ exists and is given  by 
$$\frac{\partial N_{p+s \Delta p,\xi}^{-1}}{\partial s} \Big |_{s=0}\approx \frac{N^{-1}_{p+s\Delta p,\xi} - N^{-1}_{p,\xi}}{s}.$$
\noindent In practice, the derivative may be calculated using finite-difference approximations. An analysis of the differentiability of the Monte-Carlo output at an equilibrium plasma state is depicted in Figure \ref{differentiable}.  Fig 1a visually demonstrates that the output neutral distribution varies smoothly as a function of the plasma background. If the operation is perfectly differentiable, one expects that the response to a perturbation of size $s$ to be of magnitude $O(s)$ with the constant of proportionality being precisely the derivative. Fig 1b shows this trend persists all the way down to $O(\epsilon)$ where $\epsilon$ is the machine floating point precision for the correlated Monte-Carlo output, while the response of the uncorrelated version remains $O(1)$. This precision ensures finite-difference methods converge to the expected precision \cite{finitedifferencing} of $O(\sqrt{\epsilon})$ (not depicted). With differentiability suitably ensured, the task of taking finite-difference is then relegated to the standard finite differencing methods (cf. \href{https://docs.scipy.org/doc/scipy/reference/generated/scipy.optimize.newton_krylov.html}{scipy.optimize.newton\_krylov})with a relative tolerance of $O(\sqrt{\epsilon})$.   }

\section{Simulation Setup}

The system of equations (\ref{plasma}--\ref{neutral}) is discretized on a 1-dimensional grid, assuming both rotational and poloidal symmetry. The grid spacing decreases toward the domain wall to better resolve edge effects, and upstream boundary conditions are imposed on the inner boundary. 

\subsection{Simulation Codes}
Two plasma models are implemented: first, a simplified model denoted $P_0$ for troubleshooting. This model assumes constant velocity and temperatures, and that the ion densities satisfy a second-order non-linear diffusion model. The second model, denoted $P_{1}$ is given by the UEDGE code, described in detail in \cite{Rognlien1992, Rognlien1999}. The code implements finite-volume methods for a non-linear system of coupled fluid and transport equations in the variables $p=(n_i, T_i, T_e, u_i)$ of ion density, temperature, and velocity, and electron temperature $t_e$. 

The Correlated Monte-Carlo methods are implemented using a simplified, homemade code written in Julia. This code includes 1-dimensional capabilities following large codebases such as DEGAS2 \cite{heifetz1986neutral,DEGASmanual}. Test flights are run using the attenuated absorption method and tallied with a track length estimator. 
\subsection{Coupling Methods}
Although the UEDGE code includes a sophisticated and optimized JFNK solver for time-stepping, we here opt for a more accessible setup using an external solver. Thus the above codes are used simply to call their respective functions $P_i$ and $N_{p,\xi}^{-1}$. The coupling schemes (\ref{explicitp}--\ref{explicitn}) and (\ref{implicitplasma}--\ref{implicitneutral}) are implemented in a Python overhead using SciPy's JFNK and finite-difference methods. The workflow of the implicit solver is depicted in Figure \ref{workflow} below. 

\begin{figure}[h!]
\begin{center}
    \includegraphics[width=0.45\textwidth]{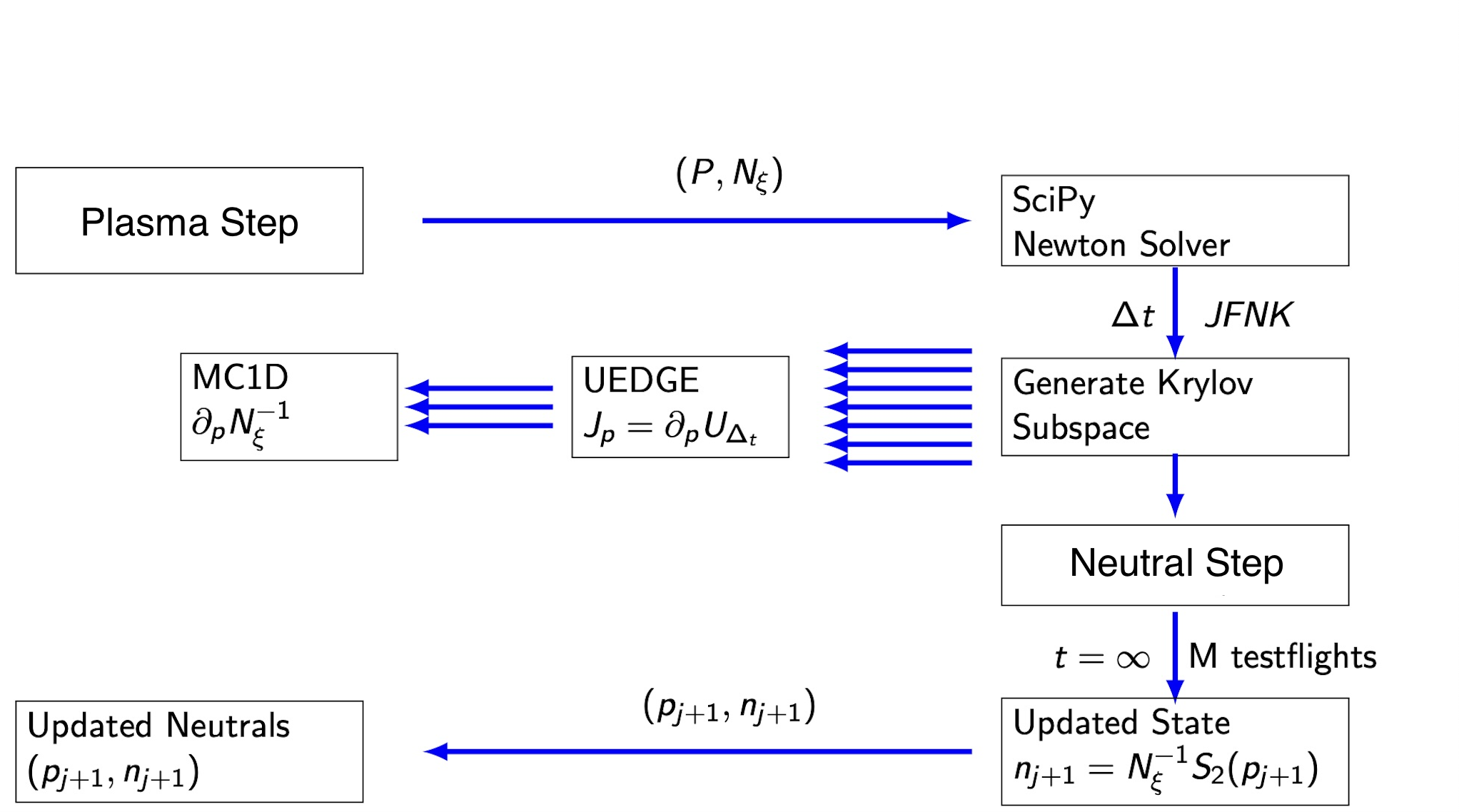}
    \end{center}
\caption{The code workflow for fully implicit coupling. The derivative of the Monte-Carlo method is called repeatedly to generate the Krylov subspace for the Jacobian during Newton iteration.}

\label{workflow}
\end{figure}

\subsection{Simulation Parameters}

Simulations are run beginning from a uniform ion density of $10^{19}$ particles/m$^2$, with uniform starting temperatures $10$ eV, and velocity $10^4$ m/s toward the divertor. A fixed value $R$ is chosen for the wall recycling coefficient (generally $R=1$ is taken). Neumann boundary conditions are imposed on the upstream plasma boundary, and Robin boundary conditions are imposed at the plate. Neutrals are assumed to have perfect reflection at walls.  

\section{Simulation results}
\label{section5}

\subsection{Correlated vs. Uncorrelated Explicit Methods}

\begin{figure}[h!]
    \centering

    \begin{subfigure}[b]{0.4\textwidth}
        \caption{}\includegraphics[width=\textwidth]{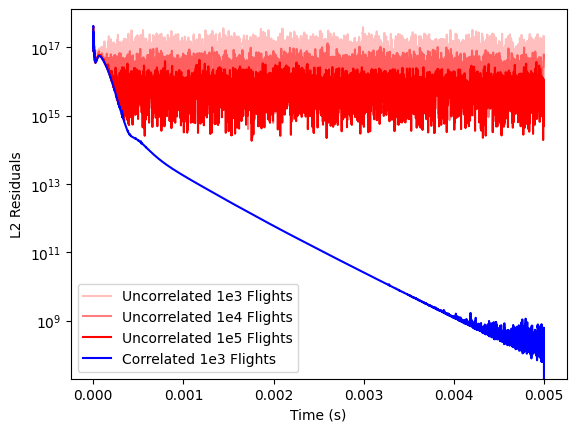}
        
        \label{fig:a}
    \end{subfigure}

    \begin{subfigure}[b]{0.4\textwidth}
        \caption{}\includegraphics[width=\textwidth]{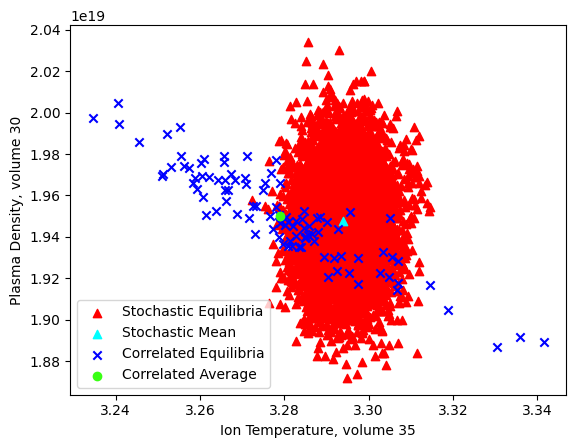}
        
        \label{fig:b}
    \end{subfigure}

    \caption{(a, top) Comparison of residual difference from converged solution for correlated vs uncorrelated explicit coupling on a log scale as a function of time. (b, bottom) The average of the stochastic equilibrium at two selected volume elements compared to the converged correlated solution for a variety of fixed random seeds.}
   \label{CMCvsMC}
\end{figure}

The convergence properties of the coupled system (\ref{plasma}--\ref{neutral}) were investigated using the explicit coupling scheme (\ref{explicitp}--\ref{explicitn}) for the UEDGE plasma model with both correlated and uncorrelated Monte-Carlo Methods. The rate of convergence is evaluated by calculating the $L^2$ residual of the plasma-neutral state, i.e. the $L^2$-norm of the difference between successive states or between the current state and a fixed reference equilibrium. The convergence of the (successive) residuals is displayed in Figure \ref{CMCvsMC} on a logarithmic scale as a function of time.

 With uncorrelated MC, the coupled system reaches a stochastic equilibrium after only a few orders of magnitude of convergence (Figure \ref{CMCvsMC}a). The magnitude of the stochastic oscillations and the amount of convergence displayed is proportional to $\sqrt{M}$ for $M$ the number of flights. In contrast, the correlated method produces convergence to the machine floating-point precision (or set tolerance of the plasma step), regardless of the number of flights. For the uncorrelated method, true equilibria are obtained by averaging the stochastic oscillations, whereas for correlated methods by averaging over different seeds (Figure \ref{CMCvsMC}b). These averages are known to each also include a bias arising from the non-linearity of the plasma model, but this is a second-order (i.e. $O(M)$) effect and becomes negligible for large particle numbers \cite{ghoos2016}, { at least in simple systems}.

\subsection{Explicit vs. Implicit Correlated Methods}

To begin, fully implicit methods are tested using a simplified plasma model described by the non-linear diffusion equation 

\begin{equation} \frac{\partial p}{\partial t}  - \nabla ( D(p)\nabla p)=0\label{simpleplasma}\end{equation}
\noindent where $D(p)=cp^n$ is the diffusion constant. The converged equilibria for the explicit and implicit coupling schemes (\ref{explicitp}--\ref{explicitn}) and (\ref{implicitplasma}--\ref{implicitneutral}) are compared below in Figure \ref{implicitvsexplicit}.

\begin{figure}[h!]
    \centering

    \begin{subfigure}[b]{0.4\textwidth}
        \caption{}\includegraphics[width=\textwidth]{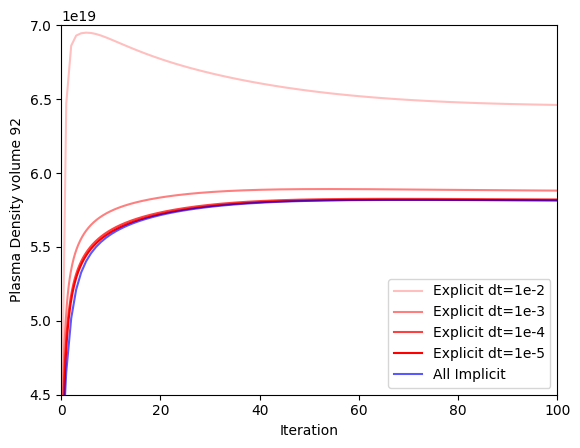}
        
        \label{fig:a}
    \end{subfigure}

    \begin{subfigure}[b]{0.4\textwidth}
        \caption{}\includegraphics[width=\textwidth]{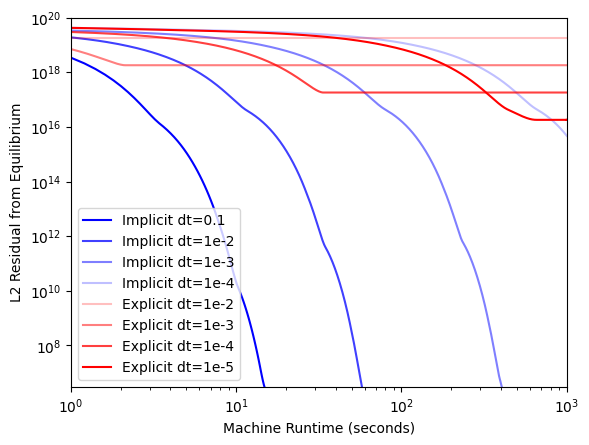}
        
        \label{fig:b}
    \end{subfigure}

    \begin{subfigure}[b]{0.4\textwidth}
        \caption{}
        \includegraphics[width=\textwidth]{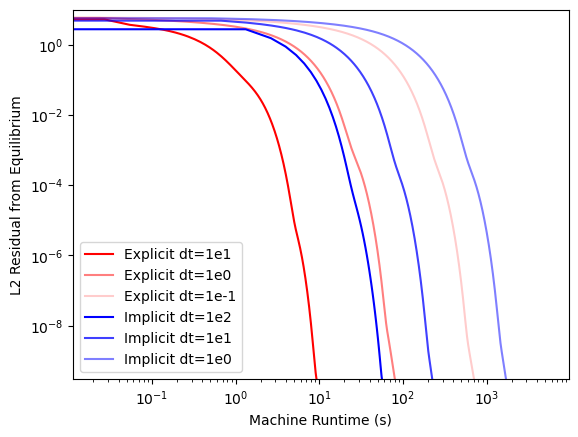}
        
        \label{fig:c}
    \end{subfigure}

    \caption{(a, top) Time evolution of plasma density at a single cell $x=3.8$ for implicit and explicit coupling with various timesteps $dt$. (b, middle) residual difference from converged solution as a function of machine runtime for n\"aive explicit and implicit methods plotted on a log-log scale. (c, bottom) The same comparison as above imposing particle conservation on the explicit method.}
   \label{implicitvsexplicit}
\end{figure}

 The implicit coupling method is found to obtain an equilibrium which is {\it independent} of the time-step used up to machine accuracy. { In contrast, the most n\"aive version of the explicit coupling method produces an equilibrium which depends on the time-step, and approaches that of the implicit method as $dt\to 0$, as depicted in Figure  \ref{implicitvsexplicit}a. } Figure \ref{implicitvsexplicit} shows the residuals with respect to the fixed timestep-independent equilibrium obtained by the implicit methods.

 Figures \ref{implicitvsexplicit}b--4c compares the residuals the two methods as a function of machine computation time. The {  n\"aive version of the} explicit coupling method is found to require exponentially more computation to reach the same levels of accuracy as the fully implicit method. The failure of a n\"aive explicit method such as this to find the correct equilibrium can be attributed to the lack of mass conservation in the explicit coupling scheme employed: the equilibria obtained by the explicit method do solve the time-independent system, but for the wrong total particle number. Indeed, with recycling coefficient $R=1$, the system obeys perfect conservation of total ion and neutral particle number; the implicit method intrinsically imposes this conservation law, while the explicit method does not. The result is that the explicit method converges to a steady-state of the system but it differs from the correct steady-state for the given initial condition insofar as the total ion and neutral particle number has an error of size $O(\Delta t)$ compared to the initial state. 
 
 {  An obvious refinement of the explicit coupling scheme is to impose conservation laws ``by hand'' by re-normalizing the total particle number after each time-step. The results of the explicit coupling scheme Eqs. \ref{explicitp}--\ref{explicitn} with this renormalization step included are displayed in Figure \ref{implicitvsexplicit}c. In this case, both methods yield the same equilibrium to machine accuracy. It should be noted that with the simple plasma model Eq. \ref{simpleplasma}, the coupled system may be solved directly with Newton iteration (i.e. taking $\Delta t\to \infty$ in either coupling scheme), thus one should be hesitant to extrapolate any conclusions (in either direction) about minimizing computation time here to more complex systems. Finally, it should be noted that hybrid implicit--explicit schemes also enjoy the advantage of obeying particle conservation without renormalization, as this is a property of the non-linear system Eqs. \ref{implicitplasma}--\ref{implicitneutral} (rather than of the solving method). }

 Of course, both the first-order (in time) explicit and implicit methods are subject to $O(\Delta t)$ numerical errors arising from discretization in the standard fashion; though the lack of dependence on $\Delta t $ for the steady-state obtained by the implicit method suggests that the constant of proportionality makes this error negligible compared to machine precision for the range of parameters explored.

\subsection{Implicit Coupling to UEDGE}
{ The fully implicit coupling scheme described here is also applicable to more sophisticated models for the plasma species (but the same model for the neutral atoms). To demonstrate this, the same coupling scheme was employed replacing the simple model (\ref{simpleplasma}) by a version of the UEDGE plasma model reduced to one spatial dimension. Convergence to machine floating-point accuracy was achieved for a variety of time steps, up to approximately $0.1$ms. In the current version of UEDGE (which incorporates a fully implicit fluid neutral model), preconditioning allows time-steps to be elongated by several orders of magnitude compared to the unconditioned system; it is expected that preconditioning could be of similar benefit here. The fully implicit model coupling UEDGE to Monte-Carlo neutral models will be further explored in future work, including investigation of convergence, preconditioning, extensions to two-dimensional systems and realistic geometries, and modeling realistic plasma scenarios. }

\section{Discussion}

{ The use of correlated Monte-Carlo methods in coupled plasma-neutral systems was investigated using both explicit and implicit coupling methods. 
Correlated methods have been investigated previously \cite{ghoos2016}, and were found to bestow little to no advantage compared to uncorrelated methods when coupling the system using explicit methods. The novel contribution of this work is to use correlation inside a Newton-Krylov solver to enable the use of fully implicit time-steps of the coupled system.

}

{ The novel implicit coupling method is demonstrated to have several notable advantages over explicit coupling schemes in the context investigated. From a naive perspective, it is not so clear that the implicit method should be advantageous; indeed, to do a single implicit time-step requires calling the Monte-Carlo solver $O(10^3)$ times to differentiate via finite-differences, generate the Krylov subspace, and perform Newton iteration. In contrast, each time-step of the explicit scheme requires only one invocation of the Monte-Carlo solver, {  and hybrid dual-stepping schemes on invocation for every dual timestep.}  The advantage of the implicit scheme appears when considering the number of calls to evolve to a particular simulation time. The finite stability regime and accuracy of explicit methods dictate that time-steps must be taken below a particular threshold. In contrast, fully implicit methods allow time-steps many orders of magnitude larger. The results in Section \ref{section5} demonstrate the ability to march towards equilibrium with large time steps retaining {  a high level of accuracy}.

The use of fully implicit solvers is not merely a theoretical exercise but has significant implications for the modeling of real-life edge plasma scenarios. On top of the simple potential for reaching equilibrium with far less computation time, fully implicit methods provide advantages for modeling the time evolution itself. Plasmas in present day devices include an overlay of physical effects spanning many different timescales. Indeed, detachment front evolution can occur on timescales of $\sim1s$ \cite{krasheninnikov1999stability}. The unconditional stability of implicit methods opens avenues for resolving these long timescale effects with time-steps of comparable size, whereas explicit (and hybrid) methods are beholden to stability constraints that force the time-step to be thousands or millions of times smaller when trying to resolve the same timescales. This stability requirement only becomes more extreme as the spatial grid is further resolved.

The robust stability and convergence properties of fully implicit methods make them ideal candidates for generalizing to study the evolution of coupled plasma-neutral systems using correlated methods in two and three-dimensional settings with realistic simulation parameters. Generalizing the methods here to realistic models incorporating more complete physics (e.g. multiple species, neutral-neutral interactions, realistic geometries and boundary conditions) presents several challenges. Most notably, the more physical complexity is included in the model, the more difficult it becomes to identify and eliminate all sources of discrete transitions to prevent ``desynchronization'' of trajectories (as discussed in Section \ref{section3.1.1}). 
 It is likely that overcoming these difficulties will require the development of more sophisticated forms of correlation, which will be the subject of forthcoming work. 
}

\section{Conclusions}

The methods investigated here display significant potential advantages over the current coupling schemes { for} fluid plasma and kinetic neutral models. First, a shift in viewpoint of Monte-Carlo methods is proposed, to one in which the code is considered as a ``black box'' producing an approximate inverse operator, which allows the use and analysis of more sophisticated (in particular implicit) coupling methods. Then, by imposing smoothness of the correlated Monte-Carlo output with respect to background plasma parameters, we show that the Monte-Carlo code may be differentiated using finite-difference schemes making it viable for use in solving algorithms based on Newton's method, in particular implicit timesteps using Krylov-based methods. {   Converged equilibria using fully implicit time evolution of coupled plasma-neutral systems are obtained using both a simplified model and the UEDGE model in one spatial dimension. The implicit method is capable, even without preconditioning, of advancing towards equilibrium with large time-steps, and shown to achieve equilibria that are independent of the time-step to machine accuracy. This opens the potential for efficient modeling of physical systems with long transients, for which explicit methods become prohibitively costly due to the necessity of small time-steps. }

\section{Acknowledgments}
G.P. is supported by an NSF Mathematical Sciences Postdoctoral Research Fellowship
(Award No. 2303102). Performed in part by LLNL under Contract DE-AC52-07NA27344. 
G.P. wishes to acknowledge Andreas Holm and George Wilkie for helpful discussions.

\bibliographystyle{apsrev4-1}
{\small 
\bibliography{mvu_psi2016_bibliography}}
\end{document}